# A signed pulse-train based image processor-array for parallel kernel convolution in vision sensors

Ahmad Reza Danesh, Mehdi Habibi

*Department of Electrical Engineering, Sensors and Interfaces Research Group, University of Isfahan, Isfahan, Iran*

## ABSTRACT

**Purpose-** High speed image processing is a challenging task for real-time applications such as product quality control of manufacturing lines. Smart image sensors use an array of in-pixel processors to facilitate high-speed real-time image processing. These sensors are usually used to perform the initial low-level bulk image filtering and enhancement.

**Design-** In this paper, using pulse-width modulated signals and regular nearest neighbor interconnections, a convolution image processor is presented. The presented processor is not only capable of processing arbitrary size kernels, but the kernel coefficients can be any arbitrary positive or negative floating number.

**Findings-** The performance of the proposed architecture is evaluated on a FPGA platform. The peak signal-to-noise ratio (PSNR) metric is used to measure the computation error for different images, filters, and illuminations. Finally, the power consumption of the circuit in different operation conditions is presented.

**Originality/Value-** The presented processor array can be used for high speed kernel convolution image processing tasks including arbitrary size edge detection and sharpening functions which require negative and fractional kernel values.

1. **INTRODUCTION**

With the advances achieved in the design of smart CMOS imagers [1, 2], many industrial and commercial measuring devices are now based on the data extracted from image sensors. The image sensor-processor combination can be used in many cases to extract the required measurements from the image data. Product quality control, 3D depth extraction, dimension measuring and robot navigation are only some examples were image sensors are used for data acquisition [3]. To boost performance and speed many recent image sensors have been equipped with in-pixel processing capabilities. Such designs can be found in image sensors which can perform data compression at the focal plane [4-6] or perform in-pixel data digitization [7, 8].

Machine vision tasks are usually algorithmically intensive and high computational power is required in the process. Moreover, the need for high speed image processing increases in applications that require real-time processing [9]. Machine vision procedures usually start-off with pixel-based image refinements, feature extractions and then narrow down to object-based interpretations of the obtained features.

Initial image processing steps usually have a high degree of parallelism and are usually concerned with the computation of a function dependent on a specific pixel value and its associated neighbors. A large group of these functions belong to the kernel-convolution processing category, which perform different types of filtering such as noise reduction, image sharpening, edge detection and etc., on the raw input image data. These steps are usually necessary in all types of machine vision algorithms to extract the required and desired features from the image for the latter stages of the machine vision task tree. Thus, parallel and array-based image processors can significantly reduce the processing time of the initial stages [10, 11, 12].

The kernel convolution function similar to many other filtering blocks uses conventional multiplication and addition steps. In most parallel implementations of the kernel convolution function such as those given in [13], [14] and [15], only processing of 3×3 kernels is possible since using adjacent links, each pixel can access the data of its neighbors only. As shown in Fig. 1(a), with a 3×3 kernel, the result of each pixel $Rs_{s,t}$ at location $(s,t)$ is obtained by:

$$Rs_{s,t} = C_{-1,-1}.P_{s-1,t-1} + C_{-1,0}.P_{s-1,t} + C_{-1,1}.P_{s-1,t+1}$$
$$+ C_{0,-1}.P_{s,t-1} + C_{0,0}.P_{s,t} + C_{+1,1}.P_{s+1,t+1}$$
$$+ C_{+1,-1}.P_{s+1,t-1} + C_{+1,0}.P_{s+1,t} + C_{+1,1}.P_{s+1,t+1} \quad (1)$$

where $C_{x,y}$ values are the kernel coefficients and $P_{s,t}$ is the pixel value at location $(s,t)$. In the kernel convolution processor presented in [16], the kernel coefficients are only limited to binary coded values.

Event driven image sensors such as [17] and [18] are able to perform large kernel size convolutions but since the result is transferred to the output using events they are more suitable for object based interpretations from the image rather than transferring of the raw filtered image to the output. Arbitrary kernel size convolution has been presented in [19] which uses an RC network expanded throughout the entire array to smooth-out the input image data. But the processor, although used to detect edges, is only capable of performing Gaussian convolutions and the kernel type is not programmable.

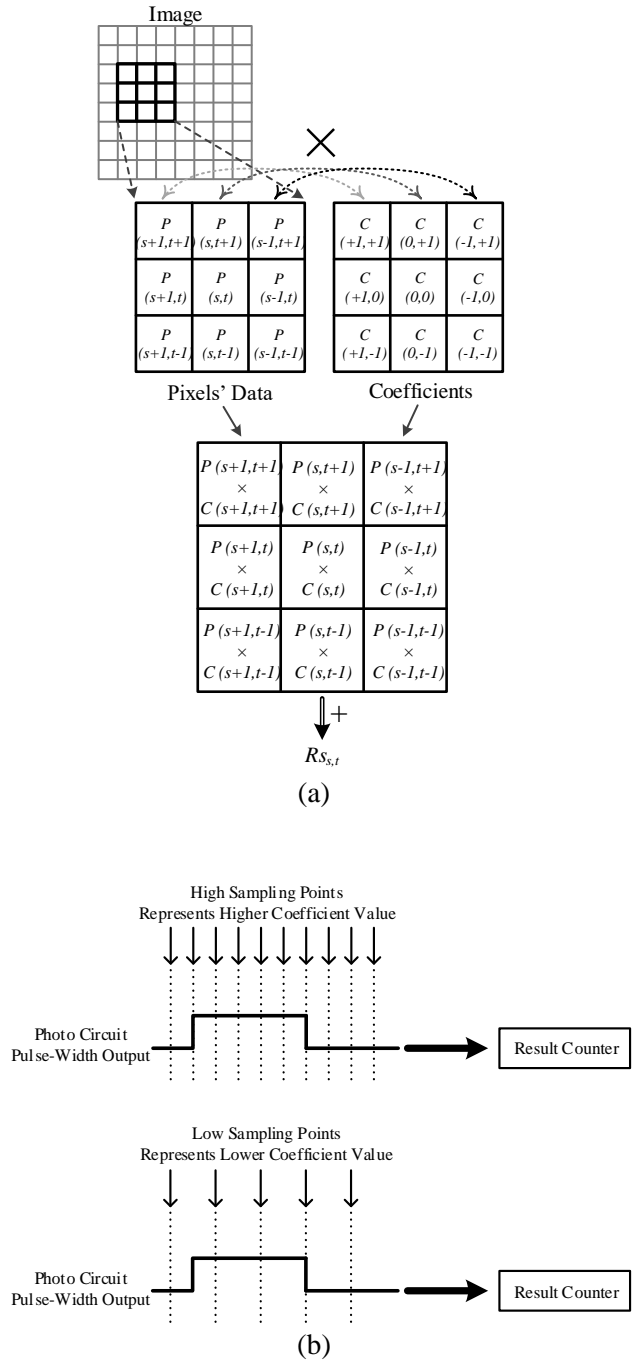

Fig. 1. a) Kernel convolution concept for a typical 3 × 3 kernel size, b) A simple view of the sampling procedure.

Arbitrary and programmable kernel convolution has been previously presented in [20]. A drawback of the method is that since it uses left shift and right shift logic for multiplication, it can only handle power of 2 coefficients. Using pulse width modulated signals, the sensor presented in [21] can perform arbitrary kernel size and arbitrary kernel value convolution. Furthermore, using pulse width processing, simplifies inter-pixel interconnects and also smaller in-pixel processors are required. However, the processor presented in [21] cannot process negative kernel coefficients. A summary of the advantages and disadvantages of the different methods can be found in Table 1.

Table 1
Comparison between advantages and disadvantages of previous kernel convolution processors.

| Work | Basic operation principle | Advantage | Disadvantage |
|---|---|---|---|
| [15, 16] | Mixed signal multiplier (binary weight coefficients and analog data) | Low element count | Kernel size limited to 3×3 |
| [17, 18] | Event based processor | Supports kernel sizes up to 32×32 elements | Outputs are based on events, not suitable for transferring entire processed image to the output |
| [19] | RC smoothing network | Arbitrary kernel size | Only Gaussian kernels are supported |
| [20] | Bit serial binary processing with inter-pixel data shifting | Arbitrary kernel size | kernel coefficients limited to power of 2 |
| [21] | PWM processor with inter-pixel sample shifting | Arbitrary kernel size and arbitrary positive kernel coefficients | negative kernel coefficients cannot be processed |

In this work using pulse width signal processing and variable frequency sampling, an arbitrary kernel value convolution processor is presented. Fig. 1(b) shows the sampling procedure concept in which the output PWM signal of each pixel is sampled according to the required coefficient and an in-pixel counter sums up the samples. With this approach, by increasing the sampling points, the counter saves a higher number. Therefore, the multiplication coefficients can be adjusted simply by changing the number of sampling points. By shifting the sampled PWM values to adjacent cells in the vertical and horizontal directions, arbitrary kernel size convolutions can be executed with the proposed processor. Furthermore, a sign bit is employed in the architecture to adjust the up or down counting of the result register in each pixel which allows the processing of signed kernel values in the convolution procedure.

With the enhancement presented for the pulse width processer, the vision processor is able to perform arbitrary kernel size, arbitrary positive and negative kernel values which are necessary for essential vision processing tasks such as edge detection. The rest of the paper is organized as follows: In Section 2, the proposed array-processor will be explained. In Section 3, the evaluation results will be presented. Finally, some concluding remarks are given in the conclusions section.

## 2. PROPOSED PROCESSOR

The proposed processor, like many other hardware that perform signal filtering, is essentially a multiply and accumulate unit and can eventually be part of an imager so that the sensor can return processed images to the output rather than raw image data. The processor has some collected features that cannot be found in previous work. These features can be summarized as follows: first, processing is commenced in parallel among all cells. That is, if $n$ clock cycles are required for the pixel processor to compute its result, within this $n$ cycle the result is completed at all pixel locations. Second, as the processing is commenced in parallel, each pixel processor is able to access data from neighbors with arbitrary distance from the target pixel. This means arbitrary kernel size convolutions can be performed with the proposed processor. Finally, the multiplication coefficients can be any arbitrary signed value.

In the proposed method, the processor of each pixel is only connected to its nearest neighbor cells. The input data feed to each pixel processor is the intensity value of the corresponding pixel and the data prepared by the nearest neighbor cells, but unlike many of the previous architectures, the input signals are neither analog type nor digital binary coded. In the proposed processor, the processing is commenced on pulse-width coded information. That is, if a pixel experiences a high illumination, its equivalent code will be a signal with a higher pulse width than that of a pixel with lower illumination level.

With this approach, every pixel data is expressed with a single digital bit line and the intensity of that pixel is expressed by the pulse width of that bit line. This technique will greatly simplify the interconnection complexity of the pixel front-end circuitry, the processor, and also the interconnections required to transfer the results of an adjacent cell to pixels further away. Furthermore, the conversion of each photodiode photocurrent-signal to a pulse-width modulated output signal, suitable for the proposed processor, is relatively straightforward and can be accomplished using a simple comparator [21]. Another aspect of the proposed architecture is that although with large kernel sizes, each pixel should have access to other pixels further away; however, with the proposed approach, interconnections from each pixel are only necessary to nearest neighbor cells. Therefore, without any direct interconnections to cells further away, the processing can be commenced by shifting data from one cell to another in each step, so that each pixel can have access to data further away without the need for any additional interconnects. Similar to [20] and [21], the 2D kernel-convolution process is accomplished by two consecutive 1D convolutions. Application of a coefficient to a specific pulse-width modulated signal is accomplished by adjusting the rate at which the signal is sampled. The final summation is achieved by counting the number of sampled pulses. It should be noted that after the completion of the first 1D convolution computation phase, the result is stored in the internal pixel counter in binary format. For the second stage 1D convolution, the binary result is converted to a single bit pulse width output by counting back (or forward based on the coefficient sign) until the counter register reaches zero. The new generated pulse width signal is used as the input for the second 1D convolution procedure while the final computation result is generated as a binary value in the final result register. In this work, by adjusting the up or down counting direction of the counting register in each phase, negative kernel values can also be processed for image filtering.

The functional behavior of the proposed processor can be described with two consecutive 1D vertical and horizontal convolutions. With this approach, the 2D kernel convolution is expressed as follows:

$$Rs_{s,t} = (\boldsymbol{P}_{s,t} \times \begin{bmatrix} \vdots \\ Cm_{j-1} \\ Cm_j \\ Cm_{j+1} \\ \vdots \end{bmatrix})^T \times [\cdots Cn_{i-1} \quad Cn_i \quad Cn_{i+1} \cdots]^T \qquad (2)$$

where $Cm_y$ -$m<y<m$ are the coefficients of the $(2m+1) \times 1$ vertical kernel matrix, $Cn_x$ -$n<x<n$ are the coefficients of the $1 \times (2n+1)$ horizontal kernel matrix and $\boldsymbol{P}_{s,t}$ is a $(2n+1) \times (2m+1)$ matrix built from the target pixel value at location $(s,t)$ at its center and the neighbors of that pixel $2m$ pixels horizontal and $2n$ pixels vertical, away in the corresponding indices. The resulting scalar value $Rs_{s,t}$ represents the output value of the processor for pixel at location $(s,t)$.

In the vertical step, the process can be simplified to the following multiply and accumulate equation:

$$Rt_{s,t} = \sum_{x=-m}^{m} Cm_x \cdot P_{s,t+x} \qquad (3)$$

With the first step completed, the initial pixel value $P_{s,t}$, will be substituted with the result value $Rt_{s,t}$. This updated value of $P_{s,t}$ will be used in the second step convolution as follows:

$$Rs_{s,t} = \sum_{y=-n}^{n} Cn_y . Rt_{s+y,t} \qquad (4)$$

In (4), $Rs_{s,t}$ will be the final result. It should be noted again that this procedure is commenced throughout the array in parallel.

Considering the vertical convolution step for brevity, the architecture presented here uses the duty cycle of a PWM signal to express the current pixel value $P_{s,t}$, which is notated by $DT\_P_{s,t}$. A frequency modulated signal notated by $f\_Cm$ is also used to express $Cm$ coefficients. The duty cycle of the PWM signal is used to gate the frequency modulated signal. Since the structure works by counting the number of cycles in the gated frequency signal during the counting interval $t_{count}$, the number of pulses in the frequency modulated signal after a duration of $t_{count}$ is given by $f\_Cm.t_{count}$. Thus the number of gated cycles during $t_{count}$ can be expressed as follows:

$$f\_Cm.t_{count}.DT\_P_{s,t} \qquad (5)$$

Now if for certain coefficients, the result register is counted backwards, the total number of counted pulses will be equal to

$$(f\_Cm^+ - f\_Cm^-).t_{count}.DT\_P_{s,t} \qquad (6)$$

where $f\_Cm^+$ is the frequency of the positive coefficient and $f\_Cm^-$ is the frequency of the negative coefficient. With this approach $(f\_Cm^+ - f\_Cm^-)$ can represent a signed coefficient of the corresponding convolution kernel and will be notated by $fs\_Cm$.

Finally, the entire counting is performed on a single result register inside each pixel. In this case, by shifting the sampled PWM signal to the adjacent neighbor cell and by adjusting the corresponding global coefficient clock signal frequency at each iteration, the total number of counted pulses in each pixel will be equal to:

$$Rt_{s,t} = t_{count} \sum_{i=-m}^{m} fs\_Cm_i \cdot (DT\_P_{s,t})_i \qquad (7)$$

where *i* refers to the iteration cycle and as explained earlier $fs\_Cm_i$ can acquire both positive and negative values which allows signed kernel convolution processing. The obtained equation resembles the required function for the kernel convolution step extracted in (4). The same procedure applies for the horizontal convolution step.

With the obtained equations, both negative and positive coefficients can be handled by the hardware with adjusting $f\_Cm^+$ and $f\_Cm^-$. As can be seen from the obtained equation, with the proposed architecture, convolution can be performed without using binary multipliers which consume considerable amount of area and interconnects. In the proposed method with just a few bit lines between adjacent cells which shift the sampled PWM signal to adjacent cells, the entire 2D convolution can be completed in parallel throughout the array.

The digital processor circuit is demonstrated in Fig. 2. The 'Sampler' unit samples the input PWM signal and the 'Data-Shifter' block sends the samples to designated neighbor pixels. Moreover, the 'Sign-Generator' unit produces the computation sign of the pixel depending on the global sign status and the related coefficient's sign. 'Sign-Shifter' unit also shifts the sign status of the pixel to its neighbors. Finally, the 'Counter' unit stores the final result in binary format.

The hardware architecture of three adjacent cells for the vertical convolution stage is shown in Fig. 3. The processing is initiated by sampling the PWM signal generated by the photodiode circuitry using the coefficient signal clock. Each sampled bit is then shifted one-step up and one-step down after each counting phase is completed, so that every cell can have access to this sampled pixel value. When this sample reaches each neighbor, if it is a logic high, the pixel's internal counter is increased based on the pixel clock counter.

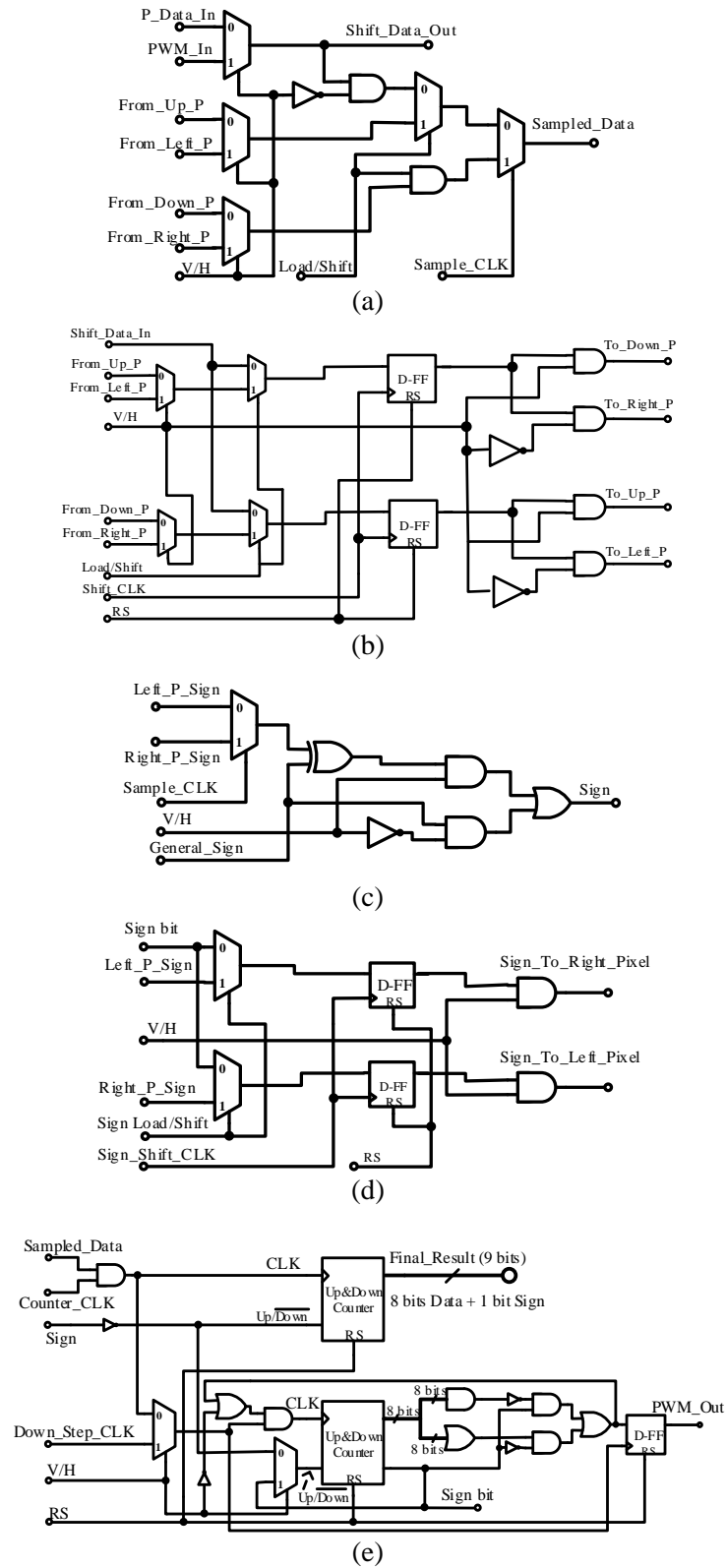

Fig. 2. Proposed processor circuit. a) Sampler, b) Data-Shifter, c) Sign-Generator, d) Sign-Shifter, and e) Result Counter

With the proposed processor, higher clock frequencies represent higher coefficient values and lower frequencies represent lower coefficients. In this work, a sign bit is also used to allow the internal counter to count backwards for negative coefficients.

The hardware configuration of three adjacent cells in the horizontal convolution phase is also shown in Fig. 4. The procedure is very similar to the vertical step except the input PWM signal is not received from the photodiode circuitry but rather from the internal counter-register's result in the previous vertical convolution step. For this purpose, the internal pixel counter-register is counted backwards until it reaches zero (if the current value is positive) so that a new PWM signal is generated for the horizontal convolution step.

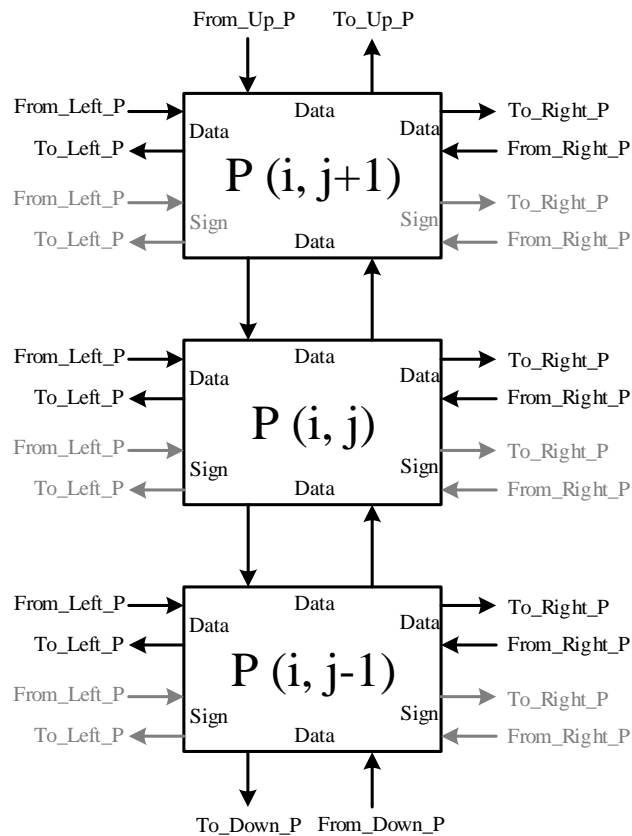

Fig. 3. Vertical configuration of three adjacent cells. The dark gray and light gray lines show data and sign interconnections respectively.

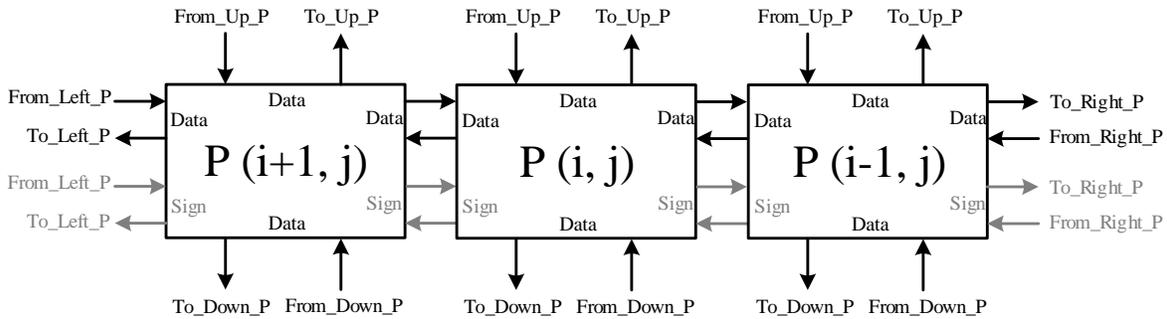

Fig. 4. Horizontal configuration of three adjacent cells. The dark gray and light gray lines show data and sign interconnections respectively.

## 3. EVALUATION RESULTS

The proposed time-domain signed pulse-width sampling approach was evaluated for filter kernels with signed coefficients. The evaluated kernels include two different edge-extraction filter types, a 5×5 Laplacian of Gaussian (LOG) and a sharpening kernel. Table 2 shows the kernel matrixes for these operations. The supporting images are obtained by behavioral simulation of the proposed architecture. All other data such as power usage and speed are given based on the design software reports when the processor is synthesized on a Virtex 7 platform.

Table 2

Common kernel matrixes with negative coefficient values

| Kernel | Matrix Array |
|---|---|
| Edge Detection-1 | $\begin{bmatrix} -1 & -1 & -1 \\ -1 & 8 & -1 \\ -1 & -1 & -1 \end{bmatrix}$ |
| Edge Detection-2 | $\begin{bmatrix} 1 & 0 & -1 \\ 0 & 0 & 0 \\ -1 & 0 & 1 \end{bmatrix}$ |
| Laplacian of Gaussian | $\begin{bmatrix} 0 & 0 & 1 & 0 & 0 \\ 0 & 1 & 2 & 1 & 0 \\ 1 & 2 & -16 & 2 & 1 \\ 0 & 1 & 2 & 1 & 0 \\ 0 & 0 & 1 & 0 & 0 \end{bmatrix}$ |
| Sharpening | $\begin{bmatrix} 0 & -1 & 0 \\ -1 & 5 & -1 \\ 0 & -1 & 0 \end{bmatrix}$ |

The evaluations were performed under different light level conditions to show the capabilities of the proposed processor on normalizing the applied kernels using processing sampling rate adjustment. The normalization functionality essentially means that the technique is able to process arbitrary kernel values. To evaluate the performance of the presented processor, it is synthesized and evaluated on a Virtex-7 platform and the power dissipation of the processor is extracted for different sampling rates. The results are shown in Table 3. As it can be inferred from the table, by increasing the sampling rate, the power consumption of the processor is raised, which is due to the increase of the system clock frequency.

Table 3

Power consumption in different sample rates, frequencies, and frame rates for one pixel-processor

| Kernel (3 × 3) | | | | | |
| --- | --- | --- | --- | --- | --- |
| Frame Rate = 1 KHz | | | Frame Rate = 10 KHz | | |
| Sample Rate | Clock (MHz) | Power ($\mu$W) | Sample Rate | Clock (MHz) | Power ($\mu$W) |
| 64 | 0.768 | ≤ 15 | 64 | 7.680 | 115 |
| 128 | 1.536 | 22 | 128 | 15.360 | 205 |
| 192 | 2.304 | 38 | 192 | 23.040 | 288 |
| 256 | 3.072 | 46 | 256 | 30.720 | 422 |

Due to the limited sampling frequency and also the use of 8-bit wide result registers, the presented approach suffers from certain error in comparison with the ideal case. To measure the computation error, the peak signal-to-noise ratio (PSNR) metric is used. The PSNR evaluates the ratio of the maximum feasible power of a signal to the power of the noise. It must be noticed that in this case, noise is the error, which is caused due to limited in-pixel storage register resolution and consequently throwing away some of the data in each calculation step. The PSNR function for 8-bit resolution registers and 1-bit sign is as follows [22]:

$$PSNR(I, A) = 20 \log_{10} 255 - 10 \log_{10} MSE(I, A) \qquad (8)$$

where, $I$ and $A$ refer to the Ideal and Actual results respectively and $MSE\ (I,A)$ is the mean square error, which can be achieved as follows [22]:

$$MSE\ (I,A) = \frac{1}{i \times j} \sum_{i,j=-k}^{k}(I(i,j) - A(i,j))^2 \qquad (9)$$

According to the mentioned equations, if the error between the Ideal and Actual results is small and MSE reaches zero, the value of the PSNR goes to infinity [22]. Therefore, a higher PSNR value shows higher quality in the processed result. The PSNR results of the presented processor for different images (shown in Appendix A section) with diverse conditions in illumination and sampling rate are provided in Table 4.

Based on the results, for low-illumination images, the PSNR value reaches infinity (no difference between the ideal and actual computation results) using a high-frequency clock pulse for some kernels that have small coefficients while for other kernels such as Laplacian of Gaussian that has large coefficients, the maximum value of PSNR is finite. In a special case for Edge Detection-2 matrix, even for normal-illumination images, the PSNR value is infinity when the clock frequency is high because the coefficients of the matrix are small enough, and also the summation of all arrays in each raw or column is zero.

The results show that the proposed processor is capable of processing kernel convolution on different images with various illuminations. It is important to note that while computation error grows in conventional digital processors with low-illumination images due to quantization noise, however, in the proposed technique, by increasing the sampling rate and thus applying a higher normalization coefficient, this quantization error can be reduced.

The characteristics of previously presented convolution image sensors are compared in Table 5. As it can be seen from the table, although there are different vision processors with kernel convolution capabilities, however, the only processor, which can perform arbitrary pixel based convolutions regardless of the kernel size and the kernel values, is the architecture presented here. The processor presented here is less limited to the kernel parameters and can be used for diverse types of filtering tasks including image normalization. Although some processors such as [17] and [18] in Table 5 can process large size kernels, however, they are based on event based processing and can encounter limitation on scenes with large picture changes.

Table 4

PSNR results (dB) for various images using different sampling rates (1 kfps)

| Kernel | Image | Normal Illumination | | | Low Illumination | | |
|---|---|---|---|---|---|---|---|
| | | X1* | X2 | X3 | X1* | X2 | X3 |
| Edge Detection-1 | Baby | 27.2 | 32.5 | 36.4 | 34.5 | 40.1 | Infinity |
| | Girl | 27.2 | 33.1 | 36.9 | 34.5 | 40.2 | Infinity |
| | Moon | 26.3 | 32.1 | 35.6 | 33.6 | 39.2 | Infinity |
| Edge Detection-2 | Baby | 44.8 | 49.2 | Infinity | 46.5 | 50.5 | Infinity |
| | Girl | 45.4 | 49.9 | Infinity | 46.6 | 50.9 | Infinity |
| | Moon | 44.7 | 48.6 | Infinity | 46.2 | 49.2 | Infinity |
| Laplacian of Gaussian | Baby | 30.4 | 35.9 | 39.6 | 43.2 | 48.2 | 51.8 |
| | Girl | 30.6 | 36.5 | 40.1 | 42.5 | 48.4 | 52.2 |
| | Moon | 30.1 | 35.5 | 39.2 | 42.2 | 46.6 | 51.4 |
| Sharpening | Baby | 34.4 | 40.3 | 43.8 | 38.5 | 44.3 | Infinity |
| | Girl | 34.4 | 40.8 | 43.9 | 38.3 | 43.8 | Infinity |
| | Moon | 33.1 | 39.4 | 43.2 | 37.5 | 42.9 | Infinity |

* X1 refers to the clock frequency of 512 kHz for normal illumination and 1.024 MHz for low illumination images.

Table 5

Comparison with previous works

| Paper | [15] | [18] | [17] | [14] | [20] | [21] | **This Work** |
|---|---|---|---|---|---|---|---|
| Operation Method | Analog | Digital | Digital | Digital | Digital | PWM | PWM |
| Processing Task | Convolution | Convolution | Convolution | Edge Detection | Convolution | Convolution | Convolution |
| Max Kernel Size | 3 × 3 | 32 × 32 | 32 × 32 | 3 × 3 | Unlimited | Unlimited | Unlimited |
| Processing Speed | 10 Kfps | 1.77-20 Meps | 1.47-16.6 Meps | 30 fps | 1 Kfps | 10 Kfps | 10 Kfps |
| Clock Frequency | NA | 120 MHz | 100 MHz | NA | 70 KHz | 62 MHz | 31 MHz * |
| Output Data Format | Analog | 18-bit Digital Event | 6-bit Digital Event | Digital | 12-bit Digital | 8-bit Digital | 9-bit Digital |
| Multiplication Coefficient | Fractional | 6-bit discrete | 4-bit discrete | Fixed | Power of 2 | Fractional | Fractional |
| Signed Computation | NA | Yes | Yes | Yes | No | No | Yes |

* For a 3×3 kernel size with sample rate of 256 points per PWM cycle and frame rate of 10 Kfps

## 4. CONCLUSIONS

In this work, a kernel-convolution image sensor was presented. The processor uses pulse-width processing to allow both arbitrary size kernel and arbitrary kernel values in the convolution matrix. The kernel values are adjusted at the required value by controlling the sampling rate of coefficient clock pulse. The choice of arbitrary kernel value becomes important when the image is to be also normalized after the completion of the process. For low illumination images, larger kernel values can be used to produce a normalized output image result while for higher illuminations, lower kernel values can be applied. It was also shown that although pulse-width processing is used in the approach, however negative kernels can be supported by the proposed method using a sign bit, which determines whether the result register counting the number of PWM samples should count up or down. The evaluation results show the effectiveness of the approach in the processing of many essential and important machine vision steps such as sharpening, edge detections, and Laplacian filter. The proposed hardware is suitable to be used in smart vision sensors for real time, high-speed and parallel image processing scenarios such as product line quality control [23, 24] and robot navigation [25, 26].

# REFERENCES

[1] Gouveia, L.C.P. and Choubey, B. (2016), "Advances on CMOS image sensors", *Emerald, J. Sensor Review,* Vol. 36 No. 3, pp. 231-239.

[2] Bloss, R. (2017), "Latest in VISION SENSOR technology as well as innovations in sensing, pressure, force, medical, particle size and many other applications", *Emerald, J. Sensor Review,* Vol. 37 No. 1, pp. 7-11.

[3] Bogue, R. (2017), "Sensors key to advances in precision agriculture", *Emerald, J. Sensor Review,* Vol. 37 No. 1, pp. 1-6.

[4] M. Zhang and A. Bermak, "Compressive Acquisition CMOS Image Sensor: From the Algorithm to Hardware Implementation," *IEEE Transactions on Very Large Scale Integration (VLSI) Systems*, vol. 18, no. 3, pp. 490-500, March 2010.

[5] S. Chen, A. Bermak and Y. Wang, "A CMOS Image Sensor with On-Chip Image Compression Based on Predictive Boundary Adaptation and Memoryless QTD Algorithm," *IEEE Transactions on Very Large Scale Integration (VLSI) Systems*, vol. 19, no. 4, pp. 538-547, April 2011.

[6] Z. Huang *et al.*, "A Vector-Quantization Compression Circuit with On-Chip Learning Ability for High-Speed Image Sensor," *IEEE Access*, vol. 5, pp. 22132-22143, 2017.

[7] A. Bermak and Yat-Fong Yung, "A DPS array with programmable resolution and reconfigurable conversion time," *IEEE Transactions on Very Large Scale Integration (VLSI) Systems*, vol. 14, no. 1, pp. 15-22, Jan. 2006.

[8] Y. Chen, F. Yuan, G. Khan, "A wide dynamic range CMOS image sensor with pulse-frequency-modulation and in-pixel amplification," *Microelectronics J.*, vol. 40, issue 10, pp. 1496-1501, 2009.

[9] M. Bigas, E. Cabruja, J. Forest, J. Salvi, "Review of CMOS image sensors," *Microelectronics J.*, vol. 37, issue 5, pp. 433-451, 2006.

[10] C. Fan, F. Li, X. Cao, B. Qian, P. Song, "A parallel arithmetic for hardware realization of digital filters," *Microelectronics J.*, vol. 83, pp. 131-136, 2019.

[11] A. Ibrahim, "Scalable digit-serial processor array architecture for finite field division," *Microelectronics J.*, vol. 85, pp. 83-91, 2019.

[12] A. Vasudevan, A. Anderson and D. Gregg, "Parallel Multi Channel convolution using General Matrix Multiplication," *2017 IEEE 28th International Conference on Application-specific Systems, Architectures and Processors (ASAP)*, Seattle, WA, 2017, pp. 19-24.

[13] F. Fons, M. Fons, E. Canto, "Run-time self-reconfigurable 2D convolver for adaptive image processing," *Microelectronics J.*, vol. 42, issue 1, pp. 204-217, 2011.

[14] C. Lee, W. Chao, S. Lee, J. Hone, A. Molnar and S. H. Hong, "A Low-Power Edge Detection Image Sensor Based on Parallel Digital Pulse Computation," *IEEE Transactions on Circuits and Systems II: Express Briefs*, vol. 62, no. 11, pp. 1043-1047, Nov. 2015.

[15] K. Ito, M. Ogawa and T. Shibata, "A variable-kernel flash-convolution image filtering processor," *2003 IEEE International Solid-State Circuits Conference, 2003. Digest of Technical Papers. ISSCC.*, San Francisco, CA, USA, 2003, pp. 470-508 vol.1.

[16] J. Dubois, D. Ginhac, M. Paindavoine and B. Heyrman, "A 10 000 fps CMOS Sensor with Massively Parallel Image Processing," *IEEE Journal of Solid-State Circuits*, vol. 43, no. 3, pp. 706-717, March 2008.

[17] L. Camunas-Mesa, C. Zamarreno-Ramos, A. Linares-Barranco, A. J. Acosta-Jimenez, T. Serrano-Gotarredona and B. Linares-Barranco, "An Event-Driven Multi-Kernel Convolution Processor Module for Event-Driven Vision Sensors," *IEEE Journal of Solid-State Circuits*, vol. 47, no. 2, pp. 504-517, Feb. 2012.

[18] L. Camunas-Mesa, A. Acosta-Jimenez, C. Zamarreno-Ramos, T. Serrano-Gotarredona and B. Linares-Barranco, "A 32×32 Pixel Convolution Processor Chip for Address Event Vision Sensors With 155 ns Event Latency and 20 Meps Throughput," *IEEE Transactions on Circuits and Systems I: Regular Papers*, vol. 58, no. 4, pp. 777-790, April 2011.

[19] J. Fernandez-Berni, R. Carmona-Galan and Á. Rodriguez-Vazquez, "Ultralow-Power Processing Array for Image Enhancement and Edge Detection," *IEEE Transactions on Circuits and Systems II: Express Briefs*, vol. 59, no. 11, pp. 751-755, Nov. 2012.

[20] M, Habibi, A. Bafandeh, M. A. Montazerolghaem, "A digital array based bit serial processor for arbitrary window size kernel convolution in vision sensors," *Elsevier, J. Integration, the VLSI Journal*, vol. 47, no. 4, pp. 417-430, 2014.

[21] M. Habibi, A. R. Danesh, "A digital arbitrary size kernel convolution smart image sensor based on in-pixel pulse width processors," *Emerald, J. Sensor Review*, vol. 37, no. 4, pp. 468-477, 2017.

[22] A. Hore and D. Ziou, "Image Quality Metrics: PSNR vs. SSIM," *2010 20th International Conference on Pattern Recognition*, Istanbul, 2010, pp. 2366-2369.

[23] Kumar, B. and Ratnam, M. (2015), "Machine vision method for non-contact measurement of surface roughness of a rotating workpiece", *Emerald, J. Sensor Review*, Vol. 35, No. 1, pp. 10-19.

[24] Wei, K. Dai, Y. and Ren, B. (2019). "Automatic identification and autonomous sorting of cylindrical parts in cluttered scene based on monocular vision 3D reconstruction", *Emerald, J. Sensor Review*.

[25] Li, H. and Zhong, C. (2016), "A machine vision based autonomous navigation system for lunar rover: The model and key technique", *Emerald, J. Sensor Review*, Vol. 36 No. 4, pp. 377-385.

[26] Singh, R. and Nagla, K. S. (2019). "A modified sensor fusion framework for quantifying and removing the effect of harsh environmental condition for reliable mobile robot mapping", *Emerald, J. Sensor Review*.